\documentclass[12pt]{article}
\usepackage{amssymb}

\input{tcilatex}

\begin{document}

\bigskip\ 

\bigskip\ 

\begin{center}
\textbf{ORIENTED MATROID THEORY AS A MATHEMATICAL}

\textbf{\ }

\textbf{FRAMEWORK FOR }$\mathbf{M}$\textbf{-THEORY}

\textbf{\ }

\textbf{\ }

\smallskip\ 

J. A. Nieto\footnote{%
nieto@uas.uasnet.mx}

\smallskip

\textit{Facultad de Ciencias F\'{\i}sico-Matem\'{a}ticas de la Universidad
Aut\'{o}noma}

\textit{de Sinaloa, 80010, Culiac\'{a}n Sinaloa, M\'{e}xico}

\bigskip\ 

\bigskip\ 

\textbf{Abstract}
\end{center}

We claim that $M$(atroid) theory may provide a mathematical framework for an
underlying description of $M$-theory. Duality is the key symmetry which
motivates our proposal. The definition of an oriented matroid in terms of
the Farkas property plays a central role in our formalism. We outline how
this definition may be carried over $M$-theory. As a consequence of our
analysis we find a new type of action for extended systems which combines
dually the $p$-brane and its dual $p^{\perp }$-brane.

\bigskip\ 

\bigskip\ 

\bigskip\ 

\bigskip\ 

\bigskip\ 

Keywords: $M$-theory, matroid theory, $p$-brane physics.

Pacs numbers: 04.60.-m, 04.65.+e, 11.15.-q, 11.30.Ly

July, 2006

\newpage

In the references [1]-[5] were established a number of connections between
oriented matroid theory [6] and several ingredients of $M$-theory [7]-[9]
including $D=11$ supergravity, Chern-Simons theory, string theory, and $p$%
-brane physics. The real motivation for such connections has been to
implement a kind of "duality principle" in $M$-theory via oriented matroid
theory. As it is known duality between various superstring theories was the
key symmetry to suggest the existence of an underlying $M$-theory which
includes the five consistent superstring theories and $D=11$ supergravity.
However, in spite of several interesting proposals for describing $M$-theory
(see Ref. [10] and references therein), the precise connection between
duality and $M$-theory as well as the correct definition of $M$-theory
remains as a mystery [8]. Our claim in this work is that oriented matroid
theory may provide the necessary mathematical tools for considering a
general concept of a duality principle and consequently may establish the
bases for a definition of $M$-theory as a `duality theory'.

Let us start off with the consideration of the oriented matroid concept.
There are several equivalent definitions of an oriented matroid, but perhaps
for physicists the most convenient one is in terms of the so-called
chirotopes (see Ref. [6] and references therein), namely:

An oriented matroid $\mathcal{M}$ is a pair $(E,\chi ),$ where $E$ is a
non-empty finite set and $\chi $ (called chirotope) is a mapping $%
E^{r}\rightarrow \{-1,0,1\}$, where $r$ means the rank on $E$, satisfying
the following properties:

$(\chi 1)$ $\chi $ is not identically zero,

$(\chi 2)$ $\chi $ is alternating,

$(\chi 3)$ for all $x_{1},x_{2},...,x_{r},y_{1},y_{2},...,y_{r}\in E$ such
that

\begin{equation}
\chi (x_{1},x_{2},...,x_{r})\chi (y_{1},y_{2},...,y_{r})\neq 0,  \tag{1}
\end{equation}%
there exists an $i\in \{1,2,...,r\}$ such that

\begin{equation}
\chi (y_{i},x_{2},...,x_{r})\chi
(y_{1},y_{2},...,y_{i-1},x_{1},y_{i+1,}...,y_{r})=\chi
(x_{1},x_{2},...,x_{r})\chi (y_{1},y_{2},...,y_{r}).  \tag{2}
\end{equation}%
Here, we assume that $x_{1},x_{2},...,x_{r}\in E.$

One of the reasons why physicists may become interested in this oriented
matroid definition is because alternating objects are very well-known in
higher dimensional supergravity and $p$-branes physics (see Ref. [11] and
references therein). This can be clarified further if we consider the case
of a vector configuration in which case the chirotope $\chi $ can be written
as

\begin{equation}
\chi (\mu _{1},...,\mu _{r})\equiv sign\det (b^{\mu _{1}},...,b^{\mu
_{r}})\in \{-1,0,1\},  \tag{3}
\end{equation}%
for all $b^{\mu _{1}},...,b^{\mu _{r}}\in R^{r}$ and for all $\mu
_{1},...,\mu _{r}\in E$. In this case the expressions (1) and (2) become
connected with the Grassmann-Pl\"{u}cker relations (see Ref. [6], section
3.5). It turns out that the expression (3) can also be written in the
alternative way

\begin{equation}
\chi (\mu _{1},...,\mu _{r})\equiv sign\Sigma ^{\mu _{1}...\mu _{r}}, 
\tag{4}
\end{equation}%
where

\begin{equation}
\Sigma ^{\mu _{1}...\mu _{r}}\equiv \varepsilon
^{a_{1}...a_{r}}b_{a_{1}}^{\mu _{1}}...b_{a_{r}}^{\mu _{r}}.  \tag{5}
\end{equation}%
Here, $\varepsilon ^{a_{1}...a_{r}}$ is the completely antisymmetric symbol.
Thus, if we introduce the base $\omega _{\mu _{1}},\omega _{\mu
_{2}},...,\omega _{\mu _{r}}$ one discovers that (5) leads to the $r$-form

\begin{equation}
\mathbf{\Sigma }=\frac{1}{r!}\Sigma ^{\mu _{1}...\mu _{r}}\omega _{\mu
_{1}}\wedge \omega _{\mu _{2}}\wedge ...\wedge \omega _{\mu _{r}},  \tag{6}
\end{equation}%
which can also be written as

\begin{equation}
\mathbf{\Sigma }=\mathbf{b}_{1}\wedge \mathbf{b}_{2}\wedge ...\wedge .%
\mathbf{b}_{r}.  \tag{7}
\end{equation}%
Therefore $\mathbf{\Sigma }$ can be identified as a decomposable $r$-form
element of $\wedge _{r}R^{n}$. Here, $\wedge _{r}R^{n}$ denotes a $%
(_{r}^{n}) $-dimensional real vector space of alternating $r$-forms on $%
R^{n}.$

In Ref. [1] it was shown that a local version of (5),

\begin{equation}
\Sigma ^{\mu _{1}...\mu _{r}}\rightarrow \sigma ^{\mu _{1}...\mu _{p+1}}(\xi
),  \tag{8}
\end{equation}%
with $r=p+1$, suggested by the so-called matroid bundle [12]-[15], allows to
establish a connection between matroids and Schild type action for $p-$%
branes [16] (see also Ref. [17]),

\begin{equation}
S_{p}=\frac{1}{2}\int d^{p+1}\xi (\gamma ^{-1}\sigma ^{\mu _{1}...\mu
_{p+1}}\sigma _{\mu _{1}...\mu _{p+1}}-\gamma T_{p}^{2}),  \tag{9}
\end{equation}%
where $T_{p}$ is a fundamental constant measuring the inertia of the $p$%
-brane, $\gamma $ is a lagrange multiplier and

\begin{equation}
\sigma ^{\mu _{1}...\mu _{p+1}}=\varepsilon ^{a_{1}...a_{p+1}}v_{a_{1}}^{\mu
_{1}}(\xi )...v_{a_{p+1}}^{\mu _{p+1}}(\xi ),  \tag{10}
\end{equation}%
with

\begin{equation}
v_{a}^{\mu }(\xi )=\partial _{a}x^{\mu }(\xi ),  \tag{11}
\end{equation}%
where $x^{\mu }(\xi )$ are $d+1$-scalar fields (see Ref. [1] for details).
One of the interesting aspects of this connection is that duality becomes
part of the $p$-brane structure in a systematic way. In order to clarify
this observation let us recall the definition of the dual oriented matroid $%
\mathcal{M}^{\ast }$. It turns out that there are different equivalent forms
for introducing duality in oriented matroid theory. In terms of chirotopes
the dual oriented matroid $\mathcal{M}^{\ast }$ is defined as follows.
First, one introduces the dual chirotope $\chi ^{\ast }$ such that

\begin{equation}
\chi ^{\ast }:E^{n-r}\rightarrow \{-1,0,1\}  \tag{12}
\end{equation}%
and

\begin{equation}
(x_{1},x_{2},...,x_{n-r})\rightarrow \chi (x_{1}^{\prime },x_{2}^{\prime
},...,x_{r}^{\prime })sign(x_{1},x_{2},...,x_{n-r},x_{1}^{\prime
},x_{2}^{\prime },...,x_{r}^{\prime }),  \tag{13}
\end{equation}%
where $(x_{1}^{\prime },x_{2}^{\prime },...,x_{r}^{\prime })$ means some
permutation of $E\backslash (x_{1},x_{2},...,x_{n-r})$ and

\begin{equation}
sign(x_{1},x_{2},...,x_{n-r},x_{1}^{\prime },x_{2}^{\prime
},...,x_{r}^{\prime })  \tag{14}
\end{equation}%
is the parity of the number of inversions of $(1,2,...,n).$ It is not
difficult to see that, as in the case of ordinary matroids [18], every
oriented matroid $\mathcal{M}(E,\chi )$ has an associated unique dual $%
\mathcal{M}^{\ast }(E,\chi ^{\ast }).$ Furthermore, it is found that $%
\mathcal{M}^{\ast \ast }=\mathcal{M}$ (see Ref. [6], section 3.4). Assuming
that similar duality result should go over a matroid bundle scenario one
finds that the object $\sigma ^{\mu _{1}...\mu _{p+1}}(\xi )$ associated
with the chirotope $\chi $ of $\mathcal{M}(\xi )$ should imply an
identification of the dual of $\sigma ^{\mu _{1}...\mu _{p+1}}(\xi )$ in the
form

\begin{equation}
^{\ast }\sigma ^{\mu _{p+2}...\mu _{d+1}}=\frac{1}{(p+1)!}\varepsilon _{\mu
_{1}...\mu _{p+1}}^{\mu _{p+2}...\mu _{d+1}}\sigma ^{\mu _{1}...\mu _{p+1}},
\tag{15}
\end{equation}%
where the dual $^{\ast }\sigma ^{\mu _{p+2}...\mu _{d+1}}$ is associated
with the chirotope $\chi ^{\ast }$ of the dual oriented matroid $\mathcal{M}%
^{\ast }(\xi ).$ Here, $^{\ast }\sigma ^{\mu _{p+2}...\mu _{d+1}}$ is given
by

\begin{equation}
^{\ast }\sigma ^{\mu _{p+2}...\mu _{d+1}}=\varepsilon ^{\hat{a}_{p+2}...\hat{%
a}_{d+1}}v_{\hat{a}_{p+2}}^{\mu _{p+2}}(\xi )...v_{\hat{a}_{d+1}}^{\mu
_{d+1}}(\xi ),  \tag{16}
\end{equation}%
where

\begin{equation}
v_{\hat{a}}^{\mu }(\xi )=\partial _{\hat{a}}x^{\mu }(\xi ).  \tag{17}
\end{equation}%
In order for (16)-(17) to make sense it is necessary to consider that
locally the fiber bundle $\mathcal{E}(B,F,G)$, with base space $B,$ fiber $F$
and structural group$\ G$, is parametrized by coordinates $\xi =(\xi
^{a},\xi ^{\hat{a}}).$ Moreover, the variables $v_{a}^{\mu }(\xi )$ and $v_{%
\hat{a}}^{\mu }(\xi )$ should be associated with the horizontal part $H_{\xi
}(\mathcal{E})$ and the vertical part $V_{\xi }(\mathcal{E})$ of the tangent
bundle $T_{\xi }(\mathcal{E})$ respectively. Of course, we have that $%
v_{a}^{\mu }(\xi )$ and $v_{\hat{a}}^{\mu }(\xi )$ satisfy the orthogonality
condition

\begin{equation}
v_{\hat{a}}^{\mu }v_{a\mu }=0.  \tag{18}
\end{equation}

From the point of view of ordinary oriented matroid theory these
observations on duality must be connected with a total vector space $R^{d+1}$
corresponding to $T_{\xi }(\mathcal{E})$, a subspace $L\subseteq R^{d+1}$
corresponding to $H_{\xi }(\mathcal{E})$ and the orthogonal complement $%
L^{\perp }$ corresponding to $V_{\xi }(\mathcal{E}).$ Thus, just as $(H_{\xi
}(\mathcal{E}),V_{\xi }(\mathcal{E}))$ determine the structure of $T_{\xi }(%
\mathcal{E}),$ the dual pair $(L,$ $L^{\perp })$ determines the structure of
total space $R^{d+1}$. Therefore, it may result very convenient to be able
to introduce the concept of an oriented matroid in terms of the structure $%
(L,$ $L^{\perp })$ rather than only in terms of the subspace $L$. If this is
possible then one should be able to make a direct transition

\begin{equation}
(L,L^{\perp })\rightarrow (H_{\xi }(\mathcal{E}),V_{\xi }(\mathcal{E})), 
\tag{19}
\end{equation}%
and in this way to ensure from the beginning the duality symmetry.
Surprisingly, the mathematicians have already considered an equivalent
definition of an oriented matroid in terms of an analogue structure to the
pair $(L,L^{\perp })$. Such a definition used the concept of Farkas property
which we shall proceed now to discuss briefly (see Ref . [19] for details).

But before we address the Farkas property (Lemma) it turns out necessary to
introduce the sign vector concept. Let $E\neq \varnothing $ be a finite set.
An element $X\in \{+,-,0\}^{E}$ is called a sign vector. The positive,
negative and zero parts of $X$ are denoted by $X^{+},X^{-},X^{0}$
respectively$.$ Further, we define \textit{supp}$X\equiv X^{+}\cup X^{-}.$
Consider two sets $\mathit{S}$ and $\mathit{S}^{\prime }$ of signed vectors.
The pair $(\mathit{S},\mathit{S}^{\prime })$ is said to have the Farkas
property , if $\forall $ $e\in E$ either

$(a)$ $\exists $ $X\in \mathit{S,}$ $e\in $\textit{supp}$X$ and $X\geq 0$

\noindent or

$(b)$ $\exists $ $Y\in \mathit{S}^{\prime }\mathit{,}$ $e\in $\textit{supp}$%
Y $ and $Y\geq 0$

\noindent but not both. Here, $X\geq 0$ means that $X$ has a positive $(+)$
or a zero $(0)$ entry in every coordinate. Observe that $(\mathit{S},\mathit{%
S}^{\prime })$ has the Farkas property if and only if $(\mathit{S}^{\prime },%
\mathit{S})$ has it.

Let $\mathit{S}$ be a set of signed vectors, and let $I$ and $J$ denote
disjoint subsets of $E$. Then 
\begin{equation}
\mathit{S}\backslash I/J=\{\tilde{X}\mid \exists X\in \mathit{S,}%
X_{I}=0,X_{J}=\ast ,X=\tilde{X}\text{ on }E\backslash (I\cup J)\}  \tag{20}
\end{equation}%
is called a minor of $\mathit{S}$ (obtained by deleting $I$ and contracting $%
J$). Here, the symbol "$\ast $" denotes and arbitrary value. If $\mathit{S}$
and $\mathit{S}^{\prime }$ are sets of sign vectors on $E$, then $(\mathit{S}%
\backslash I/J,\mathit{S}^{\prime }\backslash J/I)$ is called minor of $(%
\mathit{S},\mathit{S}^{\prime })$. Similarly,

\begin{equation}
_{I}\mathit{S}=\{\tilde{X}\mid \exists X\in \mathit{S,}X_{I}=-\tilde{X}%
_{I},X_{E\backslash I}=\tilde{X}_{E\backslash I}\}  \tag{21}
\end{equation}%
is called the reorientation of $\mathit{S}$ on $I$. Further, $(_{I}\mathit{S}%
,_{I}\mathit{S}^{\prime })$ is the reorientation of $(\mathit{S},\mathit{S}%
^{\prime })$ on $I$. Moreover, $\mathit{S}$ is symmetric if $\mathit{S}=-%
\mathit{S}$, where $-\mathit{S}$ is the set of signed vectors which are
opposite to the signed vectors of $\mathit{S}$.

With these definitions at hand we are ready to give an alternative but
equivalent definition of an oriented matroid. Let $E\neq \varnothing $ be a
finite set and let $\mathit{S}$ and $\mathit{S}^{\prime }$ be two sets of
sign vectors. Then the pair $(\mathit{S},\mathit{S}^{\prime })$ is called an
oriented matroid on $E$, if [19]

$(F1)$ $\mathit{S}$ and $\mathit{S}^{\prime }$ are symmetric and

$(F2)$ every reorientation of every minor of $(\mathit{S},\mathit{S}^{\prime
})$ has the Farkas property.

\noindent From this definition follows that $(\mathit{S},\mathit{S}^{\prime
})$ is an oriented matroid if and only if $(\mathit{S}^{\prime },\mathit{S})$
is an oriented matroid and also that every minor and every reorientation of
an oriented matroid is an oriented matroid again.

Two sign vectors $X$ and $Y$ are orthogonal, denoted by $X\perp Y$, if

\begin{equation}
(X^{+}\cap Y^{+})\cup (X^{-}\cap Y^{-})\neq 0\Leftrightarrow (X^{+}\cap
Y^{-})\cup (X^{-}\cap Y^{+}).  \tag{22}
\end{equation}%
The set%
\begin{equation}
\mathit{S}^{\perp }=\{X\mid X\perp \mathit{S}\},  \tag{23}
\end{equation}%
where $X\perp \mathit{S}$ means that $X\perp Y$ for every $Y\in \mathit{S}$,
is called the orthogonal complement of $\mathit{S}$. If $\mathit{S}^{\prime
}\subseteq \mathit{S}^{\perp }$ then we say that $\mathit{S}$ and $\mathit{S}%
^{\prime }$ are orthogonal. If $(\mathit{S},\mathit{S}^{\prime })$ is an
oriented matroid then two important results follow, namely $\mathit{S}$ and $%
\mathit{S}^{\prime }$ are orthogonal and $(\mathit{S},\mathit{S}^{\perp })$
is also an oriented matroid.

A collection $\mathcal{C}$ of sign vectors on the set $E$ is the set of
signed circuits of an oriented matroid on $E$ if and only if satisfies the
following axioms:

$(C0)$ $\mathcal{C\neq \varnothing },$

$(C1)$ $\mathcal{C}$ is symmetric,

$(C2)$ for all $X$,$Y\in \mathcal{C}$, if $X\subseteq Y$, then $X=Y$ or $%
X=-Y,$

$(C3)$ for all $X$,$Y\in \mathcal{C}$, $X\neq -Y$ and $e\in X^{+}\cap Y^{-}$%
, there is a $Z$ such that $e\neq $\textit{supp}$Z$, $Z^{+}\subseteq
X^{+}\cup Y^{+}$ and $Z^{-}\subseteq X^{-}\cup Y^{-}.$

Now, it can be proved that if $(\mathit{S},\mathit{S}^{\prime })$ is an
oriented matroid then $\mathit{S}$ satisfies the axioms $(C0)-(C3)$. Indeed
it can be proved that the axioms $(F1)-(F2)$ and $(C0)-(C3)$ provide
equivalent definitions of an oriented matroid (see Ref. [19]).

The key link between the definition of an oriented matroid in terms of a
chirotope $(\chi 1)-(\chi 3)$ and circuits $(C0)-(C3)$ is provided by the
relation

\begin{equation}
\mathcal{C}(x_{i})=(-1)^{i}\chi (x_{1},...,x_{i-1},x_{i+1},...,x_{r+1}), 
\tag{24}
\end{equation}%
where $\{x_{1},...,x_{r}\}$ is a basis of $\mathcal{M}$ and

\begin{equation}
\mathcal{C\subset }\{x_{1},...,x_{r},x_{r+1}\}.  \tag{25}
\end{equation}%
Thus, although it is not straightforward to prove that the definitions $%
(\chi 1)-(\chi 3)$ and $(C0)-(C3)$ are equivalent, the relation (24) gives
an idea of the key step in such a proof. Consequently, we may conclude that $%
(\chi 1)-(\chi 3),(C0)-(C3)$ and $(F1)-(F2)$ are three equivalent
definitions of the same structure: an oriented matroid.

Now, assume that $(\mathit{S},\mathit{S}^{\perp })$ is an oriented matroid.
One question that arises is: Can $(\mathit{S},\mathit{S%
{\acute{}}%
}^{\perp })$ be realizable in terms of the pair $(L,L^{\perp })$ of dual
vector spaces, where $L,L^{\perp }\subseteq R^{d+1}$? The answer is, of
course not always. This means that the oriented matroid concept is more
general than the concept of a vector space. In fact, one can prove that a
dual pair of realizable oriented matroids on $E$ corresponds to a pair of
orthogonal subspaces in $R^{d+1}$. Moreover, one may also ask a related
question: Is it possible to connect $(L,L^{\perp })$ with a fiber bundle
structure $\mathcal{E}(B,F,G)$? Matroid bundle seems to give an affirmative
answer to this question. In fact, the main idea of a matroid bundle is to
replace the fiber $F$ by an oriented matroid $\mathcal{M}(\xi ^{a})$, where $%
\xi ^{a}$ are local coordinates on the base manifold $B$. However, this
prescription is focused more on the definitions $(\chi 1)-(\chi 3)$ and $%
(C0)-(C3)$ of an oriented matroid rather than on the definition $(F1)-(F2)$.
But considering the equivalence between the three definitions $(\chi
1)-(\chi 3),(C0)-(C3)$ and $(F1)-(F2)$ one should expect a definition for a
matroid bundle in terms of the Farkas structure $(F1)-(F2).$ Our conjecture
is that the tangent bundle $T(\mathcal{E})$ would be the central tool to
achieve this goal. The reason is that $T(\mathcal{E})$ naturally admits the
splitting

\begin{equation}
T_{\xi }(\mathcal{E})=H_{\xi }(\mathcal{E})\oplus V_{\xi }(\mathcal{E}), 
\tag{26}
\end{equation}%
in terms of the horizontal $H_{\xi }(\mathcal{E})$ and vertical $V_{\xi }(%
\mathcal{E})$ subspaces of $T(\mathcal{E})$. In fact, one may think that the
object $(H_{\xi }(\mathcal{E}),V_{\xi }(\mathcal{E}))$ constitutes a local
version of the pair of complementary subspaces $(L,L^{\perp })$. Therefore,
just as the matroid bundle concept considers the transition $\mathcal{M}%
\rightarrow \mathcal{M}(\xi ^{a})$, based on any of the definitions $(\chi
1)-(\chi 3)$ and $(C0)-(C3),$ one may assume that equivalent definition for
a matroid bundle in terms of the definition $(F1)-(F2)$ should provide the
transition $(L,L^{\perp })\rightarrow (H_{\xi }(\mathcal{E}),V_{\xi }(%
\mathcal{E}))$. For physicists this kind of scenario is not completely new
since in Kaluza-Klein theory through the so-called spontaneous
compactification mechanism the vector space $R^{d+1}$ (or a manifold) is
locally considered as a fiber bundle $B\times F$ and as a consequence the
tangent bundle $T_{\xi }(\mathcal{E})$ splits as in (26).

What could the connection between the matroid bundle theory in terms of the
pair $(\mathit{S},\mathit{S}^{\perp })$ and $M$-theory be? Let us assume
that the starting point in $M$-theory is a duality principle based on the
dual symmetry contained in the Farkas property $(F1)-(F2)$, then one should
expect that the pair $(\mathit{S},\mathit{S}^{\perp })$ plays a basic role
in the partition function associated with any proposed $M$-theory, which we
symbolically write as

\begin{equation}
Z_{(\mathit{S},\mathit{S}^{\perp })}=\int DX\exp (S_{(\mathit{S},\mathit{S}%
^{\perp })}).  \tag{27}
\end{equation}%
As a consequence, due to the Farkas property one ensures a duality symmetry
in (27) not only at the level of the full space $(\mathit{S},\mathit{S}%
^{\perp })$, but also for any subspace $(\mathit{S}\backslash I/J,\mathit{S}%
^{\prime }\backslash J/I)$ corresponding to any reoriented minor of $(%
\mathit{S},\mathit{S}^{\perp }).$ In other words, the partition function

\begin{equation}
Z_{(\mathit{S}\backslash I/J,\mathit{S}^{\prime }\backslash J/I)}=\int
DX\exp (S_{(\mathit{S}\backslash I/J,\mathit{S}^{\prime }\backslash J/I)}) 
\tag{28}
\end{equation}%
must also contain the dual symmetry. This suggests a split of the full
partition function $Z_{(\mathit{S},\mathit{S}^{\perp })}$ in terms of
fundamental minors $Z_{(\mathit{S}\backslash I/J,\mathit{S}^{\prime
}\backslash J/I)}$. In fact, this conclusion is closely related to the
oriented matroid result

\begin{equation}
(\mathcal{M}_{1}\oplus \mathcal{M}_{2})^{\ast }=\mathcal{M}_{1}^{\ast
}\oplus \mathcal{M}_{2}^{\ast },  \tag{29}
\end{equation}%
where $\mathcal{M}_{1}\oplus \mathcal{M}_{2}$ is the direct sum of two
oriented matroids $\mathcal{M}_{1}$ and $\mathcal{M}_{2}$ and $\mathcal{M}%
^{\ast }$ denotes the dual of $\mathcal{M}$.

In the case of $p-$branes physics one may think as follows. Let us associate
the $p_{1}-$brane and $p_{2}-$brane with the matroids $\mathcal{M}_{1}$ and $%
\mathcal{M}_{2}$ respectively, then the corresponding partition functions 
\begin{equation}
Z_{\mathcal{M}_{1}}=\int DX\exp (S_{p_{1}})  \tag{30}
\end{equation}%
and 
\begin{equation}
Z_{\mathcal{M}_{2}}=\int DX\exp (S_{p_{2}})  \tag{31}
\end{equation}%
should lead to the dual symmetry $Z=Z^{\ast }$ of the total partition
function $Z=Z_{\mathcal{M}_{2}}Z_{\mathcal{M}_{2}}.$ Here, the actions $%
S_{p_{1}}$ and $S_{p_{2}}$ are determined by (9). Due to the equivalence
between $(\chi 1)-(\chi 3)$ and $(F1)-(F2)$ one should expect that the
similar conclusion must be true using an action for $p$-branes based on the
Farkas property, which we symbolically write as $S_{(p,p^{\perp })},$ where $%
p^{\perp }=d-p-1$. But once again a question emerges: What could the form of 
$S_{(p,p^{\perp })}$ be? This action should contain the invariance $%
S_{(p,p^{\perp })}=S_{(p^{\perp },p)}$ and for that reason one should expect
that the action $S_{(p,p^{\perp })}$ would combine the objects $\sigma ^{\mu
_{1}...\mu _{p+1}}$ and $^{\ast }\sigma ^{\mu _{p+2}...\mu _{d+1}}$ which
are related according to (15). In analogy to (9) one discovers that the
simplest possibility seems to be

\begin{equation}
S_{(p,p^{\perp })}=\frac{1}{2(d+1)!}\int d^{d+1}\xi (\gamma ^{-1}\varepsilon
^{\mu _{1}...\mu _{p+1}\mu _{p+2}...\mu _{d+1}}\sigma _{\mu _{1}...\mu
_{p+1}}\text{ }^{\ast }\sigma _{\mu _{p+2}...\mu _{d+1}}-\gamma T_{p}^{2}). 
\tag{32}
\end{equation}%
We observe that the integrand in this action is over the whole fiber bundle
space $\mathcal{E}$. The reason for this is that according to (10) and (16) $%
\sigma ^{\mu _{1}...\mu _{p+1}}$ is defined over the horizontal space $%
H_{\xi }(\mathcal{E})$ while $^{\ast }\sigma _{\mu _{p+2}...\mu _{d+1}}$ is
defined over the vertical space $V_{\xi }(\mathcal{E})$ and therefore the
integrand in (32) must be over the dimension of the tangent bundle $T_{\xi }(%
\mathcal{E}),$ which is equal to $d+1$.

Using (15) it is not difficult to see that the action $S_{(p,p^{\perp })}$
can be written in any of the alternative ways:

\begin{equation}
S_{(p,0)}=\frac{1}{2(d+1)!}\int d^{d+1}\xi (\gamma ^{-1}\sigma ^{\mu
_{1}...\mu _{p+1}}\sigma _{\mu _{1}...\mu _{p+1}}-\gamma T_{p}^{2})  \tag{33}
\end{equation}%
or

\begin{equation}
S_{(0,p^{\perp })}=\frac{1}{2(d+1)!}\int d^{d+1}\xi (\gamma ^{-1\ast }\sigma
^{\mu _{p+2}...\mu _{d+1}}\text{ }^{\ast }\sigma _{\mu _{p+2}...\mu
_{d+1}}-\gamma T_{p}^{2}).  \tag{34}
\end{equation}%
We observe that in general we have $x^{\mu }(\xi ^{a},\xi ^{\hat{a}})$.
Therefore, in order to derive the action (9) from (33) we need to assume $%
x^{\mu }(\xi ^{a},\xi ^{\hat{a}})=f(\xi ^{\hat{a}})x^{\mu }(\xi ^{a})$, so
that we can integrate out the coordinates $\xi ^{\hat{a}}$. We observe ,
however, that in this case the action $S_{(0,p^{\perp })}$ vanishes.
Similarly, if we assume that $x^{\mu }(\xi ^{a},\xi ^{\hat{a}})=g(\xi
^{a})x^{\mu }(\xi ^{\hat{a}})$ then action (32) leads to a $p^{\perp }-$%
brane action, but the $p-$brane action $S_{(p,0)}$ vanishes. This means that
the action (32) is more general than the actions (33) and (34). It is
interesting to observe that in the case $d+1=2n$ the action (32) can be
reduced to the Zaikov self-dual action [20] which has been studied in some
detail by Castro [21].

In order to clarify the meaning of (32) let us assume that the action $S_{p}$
corresponds to a $1-$brane, with $d+1=8.$ In this case the dual action $%
S_{p^{\perp }}$ corresponds to a $5-$brane action. But, the interesting
aspect of (32) is that $S_{(p,p^{\perp })}$ combines both the $1-$brane and $%
5-$brane in a unified way. Thus, the corresponding partition function 
\begin{equation}
Z_{(\mathit{S},\mathit{S}^{\perp })}=\int DX\exp (S_{(1,5)}(\mathit{S},%
\mathit{S}^{\perp }))  \tag{35}
\end{equation}%
should describe the quantum dynamics of both the $1-$brane and $5-$brane.

Let us make some final comments. Just as duality is a concept of fundamental
importance in $M$-theory the same is true for oriented matroid theory. In
fact, all objects defined in oriented matroid theory can be dualized. In
particular since $\mathcal{M}^{\ast \ast }=\mathcal{M},$ an oriented matroid 
$\mathcal{M}$ can be defined either directly or dually. Due to this duality
features of oriented matroid theory, perhaps a more appropriate name for
matroid theory should be `duality theory'. In view of the duality definition
(12)-(14) in terms of chirotopes it becomes evident that oriented matroid
may provide the mathematical framework for implementing a duality principle
in $M$-theory. This conjecture becomes more evident using the Farkas
definition of an oriented matroid. This suggests a partition function for $M$%
-theory which automatically ensures the duality symmetry not only for the
dual space ($\mathit{S},\mathit{S}^{\perp }$) but for any minor of ($\mathit{%
S},\mathit{S}^{\perp }$) as well. These observations lead us to discover the
action (32) which dually combines both the $p-$brane and its dual $p^{\perp
} $-brane.

Yet one may wonder whether the action (32) is unique. As we mentioned just
below expression (14) every oriented matroid $\mathcal{M}(E,\chi )$ has an
associated unique dual $\mathcal{M}^{\ast }(E,\chi ^{\ast })$ and $\mathcal{M%
}^{\ast \ast }=\mathcal{M}$ (see Ref. [6], section 3.4). This implies that
starting with $\sigma _{\mu _{1}...\mu _{p+1}}$ we find that the dual
chirotope $^{\ast }\sigma _{\mu _{p+2}...\mu _{d+1}}$ is unique and vice
versa starting with $^{\ast }\sigma _{\mu _{p+2}...\mu _{d+1}}$ we discover
that the chirotope $\sigma _{\mu _{1}...\mu _{p+1}}$ is unique. Thus, the
action (32) is constructed with two unique chirotopes $\sigma _{\mu
_{1}...\mu _{p+1}}$ and $^{\ast }\sigma _{\mu _{p+2}...\mu _{d+1}}$. It
remains to be seen whether the combination of $\sigma _{\mu _{1}...\mu
_{p+1}}$ and $^{\ast }\sigma _{\mu _{p+2}...\mu _{d+1}}$ given in (32) is
unique. First observe that the indices of the combination $\sigma _{\mu
_{1}...\mu _{p+1}}$ $^{\ast }\sigma _{\mu _{p+2}...\mu _{d+1}}$ go from $\mu
_{1}$ to $\mu _{d+1}$. This means that one requires a completely
antisymmetric object with $d+1$ indices. Thus, one can associate such an
object with a chirotope with $d+1$ indices. Considering a ground set of the
form $E_{d+1}=\{1,...,d+1\}$ one discovers that the object $\varepsilon
^{\mu _{1}...\mu _{p+1}\mu _{p+2}...\mu _{d+1}}$ is the required chirotope
which is unique (see Ref. [1]).

\bigskip\ 

\noindent \textbf{Acknowledgments: }I would like to thank E. Sezgin and
Texas A\&M University Physics Department for their kind hospitality, where
part of this work was developed.

\smallskip\

\end{document}